\documentclass[a4paper,11pt]{article}
\usepackage{jheppub} 
\usepackage{braket}
\usepackage{subfigure} 
\usepackage{tikz}
\usepackage{tikz-3dplot}
\usetikzlibrary{arrows.meta,calc,positioning}
\tdplotsetmaincoords{70}{120}
\tikzset{
  arr/.style={
    ->,
    >=Stealth,
    line width=0.4pt,
    shorten >=1pt,
    shorten <=1pt
  }
}
\hypersetup{
    colorlinks=true,
    linkcolor=magenta,
    citecolor=blue,
    urlcolor=cyan
}

\newcommand\eea{\end{eqnarray}}
\newcommand\bea{\begin{eqnarray}}
\newcommand{\be}{\begin{equation}}
\newcommand{\ee}{\end{equation}}
\newcommand{\Tr}{\text{Tr}}
\newcommand{\ThetaF}[2]{\Theta\left[ \begin{array}{c} #1 \\ #2  \end{array} \right]}
\newcommand{\x}{{\xi}}
\newcommand{\alg}{\mathcal{A}}

\title{Universality of Magic in Local Quantum Field Theory}

 \author[a,b]{Valentin Benedetti}
 \author[a]{Atish Dabholkar}
 \author[a,c,d]{Marcello Dalmonte}
 \affiliation[a]{The Abdus Salam International Centre for Theoretical Physics, \\Strada Costiera 11, Trieste 34151, Italy}
 \affiliation[b]{INFN, Sezione di Trieste,\\ Via Valerio 2, I-34127 Trieste, Italy}
  \affiliation[c]{INFN, Sezione di Bologna,\\ Via Irnerio 46, I-40126 Bologna, Italy}
  \affiliation[d]{Dipartimento di Fisica e Astronomia, Università di Bologna,\\ Via Irnerio 46, I-40126 Bologna, Italy}

\abstract{We show that no stabilizer state in a discrete realization of a local quantum field theory can flow in the continuum to the vacuum or to any state that resembles the vacuum at short distances. The argument rests on the fact that the entanglement spectrum is flat for stabilizer states but non-flat for cyclic and separating states in a local QFT as a consequence of the type III$_1$ nature of the von Neumann algebras associated with arbitrary subregions. Our result implies that simulating physically relevant QFT states necessarily requires quantum resources beyond stabilizer states and Clifford operations. We comment on the implications for holography.}

\begin{document}
\vspace*{1.5cm}
\maketitle
\flushbottom

\section{Introduction}
\label{sec:intro}

Stabilizer states and Clifford operations are the ones which are easier to implement physically within the framework of quantum error correction based on stabilizer codes~\cite{nielsen2010quantum}. By virtue of the Gottesman-Knill theorem \cite{Gottesman:1997zz,Gottesman:1998hu}, results of quantum computations performed only using Clifford operations acting upon stabilizer states can also be obtained on a classical computer in polynomial time \cite{Aaronson:2004xuh}. Hence, such computations do not offer a quantum advantage. 
 
There is growing interest in understanding whether quantum computers and simulators can simulate quantum field theory phenomena, in particular in the context of gauge theories~\cite{Banuls2020,bauer2023quantum}. If one wishes to simulate the vacuum of a quantum field theory on a quantum computer, it is useful to know \textit{a priori} how distinct the continuum vacuum is from the stabilizer states. This is the question we address in this note.  In the context of a discrete realization of a local\footnote{We define local to mean theories which can be defined as a net of type III$_1$ factors. For a Lagrangian QFT to be local, it is not sufficient that the Lagrangian is local; the algebra of local observables must also be non-trivial. Thus, while Yang-Mills theory is local, Chern-Simons theory or a more general topological quantum field theory is not. Our results apply to quantum field theories which are local in this sense.} quantum field theory (QFT), we show that no stabilizer state can flow to the continuum vacuum or to any state that resembles the vacuum at short distances. This very general result relies on a simple relation between `\textit{entanglement}' and `\textit{magic}', two quite distinct concepts from quantum information theory.

The study of resources that prevent a quantum state from being efficiently simulated on a classical computer is indeed a central problem in quantum resource theory \cite{Chitambar:2018rnj}. In this context, a key concept is magic \cite{magic1,magic2} 
(or non-stabilizer-ness), which quantifies how distinct an arbitrary quantum state is from a stabilizer state. Thus, magic captures the essential ingredient necessary for going beyond classical computation to achieve universal quantum computation \cite{Dawson:2005blj,magic1} that can execute arbitrary quantum algorithms. Recently, the study of magic has attracted considerable attention also in other areas of theoretical physics, including topological quantum computing and topological field theories \cite{Dennis2002,Kitaev2003}, gauge theories \cite{tarabunga2023many,Jha:2026ror}, conformal field theory \cite{Swingle:2020zoz,haug2023quantifying,Hoshino:2025jko,Hoshino:2025ine,ding2025evaluating}, and holography \cite{Cao:2024nrx,Chemissany:2025vye,Basu:2025uxw,Cao:2026uoq}.

Entanglement refers to the spooky quantum correlations that can exist in a bipartite quantum system such as a Bell pair. We emphasize that magic is a fundamentally different resource from entanglement. For instance, stabilizer states which have no magic can exhibit large - in fact, close to maximal - entanglement \cite{PRXQuantum.6.020324}, a Bell pair being the simplest example. More basically, entanglement is a form of correlation between systems (the most common one being bipartite entanglement), while magic is a property of a state, and not a mutual correlation. Nevertheless, these two resources are not entirely independent. Given a spatial bipartition that divides a system into two subsystems, pure stabilizer global states are characterized by a perfectly flat entanglement spectrum \cite{Tirrito:2023fnw,Cao:2024nrx}. In other words, all R\'enyi entropies associated with the reduced density matrix of a subsystem coincide - and, most importantly, remain so upon application of arbitrary Clifford operations on any of the partitions. 

In QFT,  entanglement is a fundamental property for any partition of a physical global pure state. This was originally established for the vacuum as a consequence of the Reeh-Schlieder theorem \cite{ReehSchlieder,WittenRev}, and extends to all states that sufficiently resemble the vacuum at short distances. Formally, this is related to the fact that algebras associated with regions in QFT are generically of type III$_1$ \cite{Fredenhagen:1984dc,Buchholz:1986bg}, which enforces not only infinite entanglement but also infinite fluctuations. Nonetheless, despite its divergent nature, the study of entanglement in QFT has significantly advanced our understanding of various non-perturbative phenomena in QFT, including irreversibility theorems \cite{ir1,ir2,ir3,ir4}, energy bounds \cite{Casini:2008cr,anec1Faulkner,anec2Hartman,qnec1Faulkner,qnec2Bousso}, symmetry breaking and confinement \cite{econf,Casini:2020rgj,Ares:2022koq}, non-equilibrium physics \cite{Calabrese:2006rx,Calabrese:2020wfx}, and much more.

In this article, we aim to study magic in a local QFT.  Defining an intrinsic measure of magic in a generic QFT is challenging because it requires first identifying a computational resource on the lattice whose continuum limit is generically difficult to understand. Nevertheless, the flatness constraint~\cite{Tirrito:2023fnw} provides a necessary and resource-independent condition that can be established directly in QFT using well-known entanglement techniques. Such a constraint seems at odds with the entanglement spectra of a continuum QFT \cite{WittenRev,Chemissany:2025vye}. Indeed, as we establish, the entanglement spectrum is necessarily nonflat for all physically relevant states in quantum field theory that resemble the vacuum at sufficiently short distances. Therefore, such states in the continuum cannot be obtained as limits of discrete stabilizer states\footnote{There is a perverse possibility that a state with a flat spectrum on a lattice flows to a state with a nonflat spectrum in the continuum. We argue later that this is highly unlikely, in section \ref{sec-1}, through  mutual information, and in section \ref{sec-2}, through the spectra of the modular operators after introducing a regulator.}. To be specific, our core argument relies on the fact that the entanglement spectrum of the vacuum state is highly constrained by Lorentz symmetry through the Bisognano-Wichmann theorem \cite{BisognanoWichmann,dalmonte2022entanglement}, and such a structure cannot be reproduced by stabilizer states.

Our results imply that Lorentz symmetry and the type III$_1$ structure of local algebras impose universal constraints on the computational properties of quantum field theories. They provide an obstruction to efficient classical simulation of QFT states. Since stabilizer states constitute the maximally classically simulable sector of many-body quantum systems, the impossibility of reproducing the entanglement structure of physically relevant QFT states within the stabilizer framework implies that generic continuum quantum field theories necessarily contain non-zero magic. This notion has practical implications also for spin chains and string nets. In particular, any stabilizer states in these models that preserve such a condition upon scaling of the net cannot correspond in the continuum to a physical state in a QFT with local degrees of freedom.

Note that it is possible to construct states in a local quantum field theory that have a flat entanglement spectrum, at least for some partitions. The simplest example is the Rindler vacuum in Rindler-Minkowski spacetime \cite{Unruh:1976db} or the Boulware vacuum in black hole spacetime \cite{Boulware:1974dm}. These states artificially remove the entanglement between the left and right wedges by construction. Thus, they can have a flat entanglement spectrum if one chooses the left-right bipartition\footnote{Although the spectrum in general will not be flat if one chooses a different bipartition.}. Our results imply that stabilizer states in the discrete version could at best simulate states in the continuum similar to the Rindler or Boulware vacuum, which are not cyclic and separating.  However, it is well-known that such states are not physically sensible. For example, the expectation value of the renormalized energy momentum tensor in the Rindler or the Boulware vacuum diverges near the horizon. Therefore, if the theory is coupled to gravity, the gravitational backreaction diverges near the horizon invalidating the semiclassical analysis \cite{Birrell:1982ix}. 

The outline of this article is as follows. In section \ref{sec-1}, we review the definition of stabilizer states in arbitrary qudit chains and their consequences for the entanglement spectrum. In section \ref{sec-2}, we show that the standard entanglement spectrum in QFT for states that resemble the vacuum at short distances does not allow for a flat spectrum. This implies that such states cannot be stabilizer states and therefore necessarily possess magic. In section \ref{sec-3}, we study several examples, including the vacuum state of various conformal field theories in different dimensions and particle states in free field theory. We conclude in section \ref{conclusions} with comments on implications for gravity and holography.

\section{Qudit stabilizer states \label{sec-1}}

In this section, we review the basics of stabilizer states in qudit spin chains. In particular, in section \ref{sec-1A} we define stabilizer states while also introducing, as a warm-up, some basic aspects of type I von Neumann algebras. In section \ref{sec-1B}, we present the standard argument showing that the entanglement spectrum of stabilizer states is necessarily flat. We also provide some elementary examples involving one and two-qubit systems.

\subsection{Definitions and properties in qudit chains \label{sec-1A}}

Our starting setup will be a qudit chain of $L$ sites describing a Hilbert space $\mathcal{H}_{d,L}$ of dimension $d^L$. Generally, the full operator algebra of this theory is a type I$_n$ factor. These are defined as von Neumann algebras without a center that are isomorphic to the space of bounded operators  $\mathcal{B}(\mathcal{H})$ on some Hilbert space obeying $\dim(\mathcal{H}) = n$, which in our case of interest is $\mathcal{H}_{d,L}$. By definition, all type I$_n$ factors are isomorphic to each other, and to the space of $n \times n$ matrices $\mathcal{M}_n(\mathbb{C})$, so we are allowed to pick any Hilbert space $\mathcal{H}$ of the appropriate dimension to represent operators and states. For us, the qudit chain picture with $n=d^L$ for $d$ prime will give us a suitable geometric and computational interpretation that will be useful later.

For a single qudit, with allowed  states $\left| j \right>$ for $j=1,2,\dots,d $,  every operator can be written as a linear combination of arbitrary powers of a shift operator $X$ and a phase operator $Z$ that can be defined as
\be
X = \sum_{j=0}^{d-1}\left| j + 1\text{ Mod }d\right>\left< j \right|\,,\quad Z=\sum_{j=0}^{d-1}\omega^j\left| j \right>\left< j \right|\,,
\ee
where $\omega=e^{2\pi \textrm{i} /d}$. Therefore, for a chain of $L$ sites, we can span $\mathcal{B}(\mathcal{H}_{d,L})$ by using the $L$-qudit string \cite{Gheorghiu:2011awf,White:2020zoz}
\be 
\mathcal{P}_{\textbf{a},\textbf{b},k}= \omega^k\bigotimes_{j=1}^L X^{a_j}Z^{b_j}\,\label{pauli-string} \,,
\ee
where $\textbf{a}$ and $\textbf{b}$  are $L$ dimensional vectors with integer entries $a_j,b_j$ taking values from $0$ to $d-1$ for all $j=1,2,\dots,L$ and $k$ is integer in the same range.
The group formed by all $\mathcal{P}_{\textbf{a},\textbf{b},k}$ and their products is called the generalized Pauli group $\mathbb{P}_{d,L}$. Note that $\mathbb{P}_{d,L}$ is different from $\mathcal{B}(\mathcal{H}_{d,L})$ as it does not include all linear combinations. 

From this perspective, stabilizer states $\left| \psi \right>\in \mathcal{H}_{d,L}$  are defined as the ones obeying the following property
\be
s \left| \psi \right> =\left| \psi \right> \,, \quad s \in \mathbb{S}_\psi\,, \label{pauli-stablizer}
\ee
for $\mathbb{S}_\psi$ the maximal non-trivial subgroup of $\mathbb{P}_{d,L}$ that leaves $\left| \psi \right> $ invariant. We emphasize that, although the description of the algebras is independent of our original choice of parametrization of the Hilbert space, the notion of stabilizerness is dependent on this choice.

The subset $\mathbb{S}_\psi$ of $\mathbb{P}_{d,L}$ describes by itself an abelian group. To show this, consider that $\mathbb{S}_\psi$ is maximal. Such a condition secures that it is indeed a group, as for $s_1,s_2 \in \mathbb{S}_\psi$ we have $ s_1 s_2\left| \psi \right> =s_1 \left| \psi \right>=\left| \psi \right>$, then $s_1 s_2 \in \mathbb{S}_\psi$. In addition, from (\ref{pauli-string}) we have that $s_1 s_2=\omega^n s_2 s_1$ for some integer $n$, and we also have from (\ref{pauli-stablizer}) that $(s_1 s_2 - s_2 s_1)\left| \psi \right>=0$. Then, necessarily $n=0$ and $s_1 s_2= s_2 s_1$ for every $s_1, s_2\in S\in \mathbb{S}_\psi$. This property allows us to write the projector onto the stabilizer subspace as 
\be
P_\psi=\frac{1}{|\mathbb{S}_\psi|}\sum_{s\in \mathbb{S}_\psi} s\,.\label{density-matrix-stablizer}
\ee
The dimension of this stabilizer subspace $\mathbb{S}_\psi$ is the same as the trace of the projector $P_\psi$. In particular, from (\ref{density-matrix-stablizer}), we have $\Tr(P_\psi)=1$. This follows from:
\be 
\Tr(\mathcal{P}_{\textbf{a},\textbf{b},k})=\begin{cases}
    \omega^k d^L & \text{if }a_j,b_j=0\,\,\forall j \\ \quad0& \text{otherwise }\,,
\end{cases}
\ee
and implies that the $P_\psi$ projects into a dimension one Hilbert space spanned by $\left| \psi \right>$. Therefore, we get that for pure stabilizer states $\psi$ the density matrix is
\be
\rho_\psi=\left| \psi \right>\left< \psi \right|=P_\psi\,.
\ee
In general, there are $d^L \prod_{k=1}^L (d^k+1)$ stabilizers states. Although they grow with system size as $d^{L^2/2 + O(L)}$, they are always a discrete set of points contained in a measure zero set of the Hilbert space. Nevertheless, stabilizers states are useful because, from (\ref{pauli-string}), they are solely determined by $k$, $\overline{a}$ and $\overline{b}$. Therefore, they can be appropriately simulated in polynomial time on a classical computer as long as we act on them with operations that preserve their stabilizer structure. These are known as Clifford operations.

To be more precise, the qudit Clifford group $\mathbb{C}_{d,L}$ is defined as the normalizer of the generalized Pauli group  $\mathbb{P}_{d,L}$ inside the unitary subgroup of bounded operators in the Hilbert space $\mathcal{H}_{d,L}$:
\be
\mathbb{C}_{d,L}\equiv \big\{ U\in \mathcal{B}(\mathcal{H}_{d,L})\,\,/\,\, UU^\dagger=1
\,\,\text{and} \,\,  U\mathcal{P}_{\textbf{a},\textbf{b},k} U^\dagger\in \mathbb{P}_{d,L}\,\, \forall\mathcal{P}_{\textbf{a},\textbf{b},k}\in \mathbb{P}_{d,L}\big\}\nonumber\,.
\ee
On each site, the resulting Clifford operators have the form \cite{Farinholt:2014wul,Hahn:2024njm}
\bea
&R& = \sum_{j,k=0}^{d-1}\omega^{jk}\left| k\right>\left< j \right|\,,\\
&P& = \sum_{j=0}^{d-1}\omega^{(j^2+j)/2} \omega^{-j(1+\text{Mod}_2[d+1])/d}\left| j \right>\left< j \right|\,,\\
&\text{SUM}&= \sum_{j,k=0}^{d-1}\big(\left| k\right>\left< k \right|+\left| k+j\right>\left< j \right|\big)\,.
\eea

\begin{figure}[t]
\centering
\tdplotsetmaincoords{65}{130}
\begin{tikzpicture}[tdplot_main_coords, scale=3,
    >=Latex,
    line cap=round,
    line join=round,
    every node/.style={font=\small}
]

  \fill[blue!10] (0,0) circle (1cm);
  \draw[very thick,->] (-1.5,0,0)--(1.6,0,0) node[below]{$x$};
  \draw[very thick,->] (0,-1.4,0)--(0,1.5,0) node[right]{$y$};
  \draw[very thick,->] (0,0,-1.25)--(0,0,1.25) node[above]{$z$};
  \draw[gray,dashed] (0,0,0) circle (1);
  \tdplotsetrotatedcoords{0}{90}{90};
  \draw[gray,dashed,tdplot_rotated_coords] (0,0,0) circle (1);
  \coordinate (Xp) at (1,0,0); \coordinate (Xm) at (-1,0,0);
  \coordinate (Yp) at (0,1,0); \coordinate (Ym) at (0,-1,0);
  \coordinate (Zp) at (0,0,1); \coordinate (Zm) at (0,0,-1);
  \draw[very thick,blue!70!black] (Zp)--(Xp)--(Yp)--(Xm)--(Ym)--cycle;
  \draw[very thick,blue!70!black] (Zp)--(Xp)--(Zm)--(Yp);
  \draw[very thick,blue!70!black] (Zp)--(Yp)--(Zm)--(Xm);
  \draw[very thick,blue!70!black] (Zp)--(Xm);
  \draw[very thick,blue!70!black] (Zp)--(Ym);
  \draw[very thick,blue!70!black] (Zm)--(Xp) (Zm)--(Yp) (Zm)--(Xm) (Zm)--(Ym);
  \draw[very thick,blue!70!black] (Ym)--(Xp);
  \begin{scope}[tdplot_screen_coords]
  \filldraw[red!70!black] (Zp) circle (0.04);
  \filldraw[red!70!black] (Zm) circle (0.04); 
  \filldraw[red!70!black] (Xp) circle (0.04); 
  \filldraw[red!70!black] (Xm) circle (0.04);
  \filldraw[red!70!black] (Yp) circle (0.04);
  \filldraw[red!70!black] (Ym) circle (0.04);
  \end{scope}
\node[red!70!black] at (0,0.2,1.05){$|0\rangle$};
\node[red!70!black] at (0,0.2,-0.95){$|1\rangle$};
\node[red!70!black] at (1,0,-0.15){$|+\rangle$};
\node[red!70!black] at (-1,0,0.15){$|-\rangle$}; 
\node[red!70!black] at (0,1,-0.15){$|i+\rangle$}; 
\node[red!70!black] at (0,-1,0.15){$|i-\rangle$}; 
\end{tikzpicture}
\caption{Stabilizer octahedron defined by the six stabilizer states on a single qubit Hilbert space inside the Bloch sphere \label{fig-bloch}}
\end{figure}
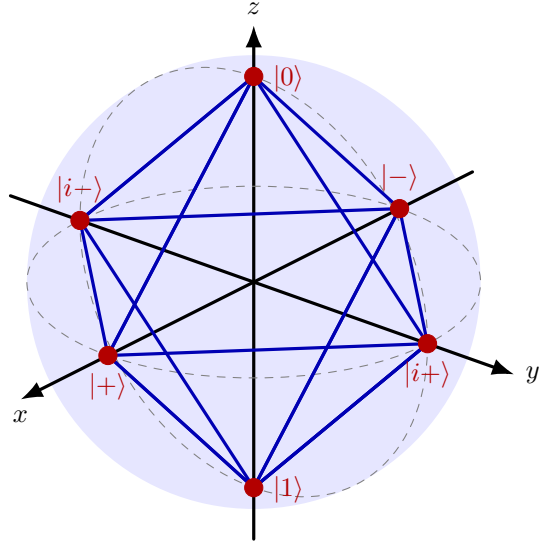

To get some intuition, let us describe the simplest example, which is a single qubit ($d=2$, $L=1$). In the Hilbert space $\mathcal{H}_{2,1}$, the space of pure states can be inscribed in the Bloch sphere as
\be 
\rho=\mathbb{I}+ v_x X + v_y Y + v_z Z, \quad v_x^2+v_y^2+v_z^2=1\,,
\ee
where $X$, $Y$ and $Z$ are the Pauli matrices defined as
\begin{align}
X &= \begin{pmatrix}
0 & 1 \\
1 & 0
\end{pmatrix}, \quad
Y = \begin{pmatrix}
0 & -i \\
i & 0
\end{pmatrix}, \quad
Z = \begin{pmatrix}
1 & 0 \\
0 & -1
\end{pmatrix}.
\end{align}
Also, $\mathbb{P}_{2,1}=\{ \pm \mathbb{I},\pm X, \pm Y,\pm Z\}$ reduces to the usual Pauli group and stabilizes six states given by the usual $|0\rangle$ and $|1\rangle$ as well as
\begin{align}
    |+\rangle &= \frac{1}{\sqrt{2}} (|0\rangle + |1\rangle),\quad |\textrm{i}+\rangle = \frac{1}{\sqrt{2}} (|0\rangle + \textrm{i}|1\rangle)\,, \\
    |-\rangle &= \frac{1}{\sqrt{2}} (|0\rangle - |1\rangle) ,\quad  |\textrm{i}-\rangle = \frac{1}{\sqrt{2}} (|0\rangle - \textrm{i}|1\rangle)\,.
\end{align}
    The Clifford group $\mathbb{C}_{2,1}$ is generated by the Pauli matrices themselves as well as the Hadamard gate $H=(X+Z)/\sqrt{2}$ that rotates around the diagonal axis situated between the $x$ and $z$ axes and the Phase gate $S=e^{\textrm{i}\pi/4}(\mathbb{I}-\textrm{i}Z)/\sqrt{2}$ that rotates around the $z$ axis. In full, they generate the 24-dimensional group that describes all possible rotations of an octahedron connecting all stabilizer points on the Bloch sphere as in figure \ref{fig-bloch}.

\subsection{Entanglement spectrum of stabilizer states \label{sec-1B}}

Following \cite{Tirrito:2023fnw}, an interesting question is how the entanglement spectrum looks for a stabilizer state. To address this, let's divide the Hilbert space as $\mathcal{H}=\mathcal{H}_A \otimes \mathcal{H}_B$, which is always possible for type I factors. In particular, let us pick for a qudit lattice a spatial partition of $L_A$ and $L_B$ sites contained inside two complementary regions $A$ and $B$, respectively, and obeying $L=L_A+L_B$. Then,  the reduced density matrix on $A$ is obtained using a projector onto the subgroup of $\mathbb{S}_\psi$ acting as the identity in $\mathcal{H}_B$:
\be 
\rho_{\psi,A}=\Tr_B(\rho_\psi)= \frac{1}{d^{L_A}}\sum_{s_A\in \mathcal{S}^A_\psi} s_A\,,
\ee
where we have factorized each string $s$ as a product $s_A\otimes s_B$. In this case, $\rho_\psi^A$ is not a projector, but it is proportional to one 
\be
\rho_{\psi,A}=\frac{1}{\lambda_A} P_{\psi,A} \,,\quad P_{\psi,A}= \frac{1}{|\mathbb{S}_\psi^A|}\sum_{s_A\in \mathbb{S}_\psi^A} s_A\,,
\ee
where $\lambda_A=d^{L_A}/{|\mathbb{S}_\psi^A|}$. The fact that $P_{\psi,A}$ is a projector implies that the eigenvalues of $\rho_{\psi,A}$ can be only $0$ or $\lambda_A^{-1} $. As a consequence, we have that the Rényi entropies exhibit a flat entanglement spectrum. In other words, the Rényi entropies $S_n(A)$ are equal for all values of the Rényi index $n$:
\be 
S_n(A)=\frac{\log\big[\Tr_A \big(\rho_{\psi,A}^n\big)\big]}{1-n}=\log\big[ \lambda_A \big]\,.\label{renyi-entropies}
\ee
Note that, although $S_n(A)$ does not depend on $n$,  it depends on the choice of bipartition, namely of $A$ and $B$; and may diverge as we take the continuum limit since $\lambda_A$ may diverge. 

Moreover, we highlight that the flatness condition is not restricted to $A$ and $B$ being topologically trivial and or connected. For example, we can take $A=A_1 \cup A_2$ and for $A_1$ and $A_2$ disconnected regions and still find Rényi entropies independent of $n$, 
\be 
S_n(A_1 \cup A_2)=\log\big[ \lambda_{A_1 \cup A_2}\big]\,.
\ee
Therefore, the flatness condition extends to the Rényi mutual informations as they are given by a sum of flat Rényi entropies of the form
\be
I_n (A_1,A_2)=S_n(A_1)+S_n(A_2)-S_n(A_1\cup A_2)\,.
\label{mutual-information-definition}
\ee
We will see in section \ref{sec-3} that this quantity is particularly useful in QFT as it is finite and independent of any choice of regulator. Therefore, its flatness should be preserved in the continuum. On the lattice, this has also been utilized in Ref.~\cite{parham2025quantumcircuitlowerbounds,korbany2025long}. 
It is important to note that, while the lack of flatness is a witness of magic, it  forms the basis for a measure of magic upon introduction of Clifford orbits (that is, the full set of operations that span the stabilizer subspace)~\cite{Tirrito:2023fnw}. It will be interesting to explore this aspect in the future, as it has not yet been developed in a QFT context. 

Another convenient quantity to discuss the QFT limit would be the Tomita modular operator. That generates an automorphism of the local algebras 
\be
\Delta_\psi^{is} \alg \Delta_\psi^{-is} =\alg\,, \quad s \in \mathbb{R}\,.
\label{modular-flow}
\ee 
In type I algebras, this is an inner automorphism generated by the density matrices
\be 
\Delta_{\psi,A} = \rho_A \otimes \rho_B^{-1}\,.
\label{modular-operator}
\ee
However, in our case, when $\psi$ is not maximally entangled, the state\footnote{The term  `state' is often used to mean a `vector' in a Hilbert space as in the case of a `stabilizer state'. In the sense of algebraic quantum field theory, a state should be understood as a positive map $\omega:\alg
\to\mathbb{C}$ obeying $\omega(\mathbb{I})=1$. A vector $\psi$   in the Hilbert space defines a state via the expectation value as $\omega(a)=\bra{\psi}a\ket{\psi}$ for all $a\in\alg$, but not all states necessarily come from vectors. See \cite{Bratteli:1979tw} for a detailed discussion. Since both are common usage,  we use `state' in both senses  depending on the context.} does not act faithfully over the sub-algebras. Namely, it has some zero eigenvalues. Therefore, we cannot compute the modular operator over the full algebras but should instead restrict it to the support of the density matrices.  The modular flow of a pure stabilizer state is trivial $\Delta_{\psi, A}=\mathbb{I}$ when acting on such support. This is because purity of the global state implies  $S_n(\rho_{\psi,A})=S_n(\rho_{\psi,B})$ forcing $\lambda_A =\lambda_B$ for complementary regions.

Finally, as an example, we consider a two-qubit system ($d=2$, $L=2$) where there are 60 stabilizer states. Of them, 36 are tensor products of the 6 qubit stabilizer states described in section \ref{sec-1A} and, therefore, are not entangled. The remaining 24 are entangled states. The simplest examples of stabilizer entangled states are given by the Bell pairs
\be 
\ket{\text{Bell}_{j,k}} = \frac{1}{\sqrt{2}}\big(\ket{0}_A \otimes\ket{j}_B + (-1)^k \ket{1}_A \otimes\ket{\overline{j}}_B\big)\,,\label{Bell-state}
\ee
for $j,k=0,1$ and $\overline{j}$ the logical negation of $j$. These can be checked case by case to be stabilized by a finite set of two-qubit Pauli strings belonging to $\mathbb{P}_{2,2}$. For example, $\ket{\text{Bell}_{0,0}}$ is stabilized by $\{I \otimes I, X \otimes X, -Y \otimes Y, Z \otimes Z\}$. 

In addition, we can check that indeed all spectra are flat. The reduced density matrix for any state of the form (\ref{Bell-state})  writes
\be
\rho_{\text{Bell}_{j,k},A} = \Tr_B \big(\ket{\text{Bell}_{j,k}}\bra{\text{Bell}_{j,k}} \big)=\frac{1}{2}\big(\ket{0}_A \bra{0}_A + \ket{1}_A \bra{1}_A \big)\,,
\ee
yielding, for all Bell states $\ket{\text{Bell}_{j,k}}$ and any Rényi index $n$, the entropies
\be 
S_n(A)=\log[2]\,.
\ee

\section{Magic in Quantum Field Theory \label{sec-2}}
In this section, we describe why physical states in QFT cannot be stabilizer states or limits thereof. In section \ref{sec-2A}, we explain why the vacuum of a generic Lorentz-invariant QFT necessarily possesses magic as a consequence of the Bisognano–Wichmann theorem. Later, in section \ref{sec-2B}, we present an argument extending this condition to all states that, at sufficiently low energies, resemble the vacuum, relating the presence of magic to the fact that algebras associated with regions in QFT are necessarily of type III$_1$.

\subsection{Magic of QFT vacuum \label{sec-2A}}

\begin{figure}[t]
\centering
\begin{tikzpicture}[scale=1.2,>=Stealth,thick]
\draw[fill=blue!10, draw=black] (-3.3,0.7) -- (0.7,0.7) -- (1,2.2) -- (-3,2.2) -- cycle;
\draw[red, thick] (-2.1,1.5) -- (-0.2,1.5);
\node[red] at (-1,1.7){\small{$A^+$}};
\draw[blue, thick] (-2.1,1.4) -- (-0.2,1.4);
\node[blue] at (-1.0,1.23){\small{$A^-$}};

\begin{scope}[shift={(0,0.3)}]
\draw[fill=blue!10, draw=black] (-3.3,-1.3) -- (0.7,-1.3) -- (1,0.2) -- (-3,0.2) -- cycle;
\draw[blue, thick] (-2.1,-0.5) -- (-0.2,-0.5);
\node[blue] at (-1,-0.3){\small{$A^+$}};
\draw[teal, thick] (-2.1,-0.6) -- (-0.2,-0.6);
\node[teal] at (-1.0,-0.77){\small{$A^-$}};
\end{scope}

\begin{scope}[shift={(-5,-3.4)}]
\draw[fill=blue!10, draw=black] (1.7,0.7) -- (5.7,0.7) -- (6,2.2) -- (2,2.2) -- cycle;
\draw[teal, thick] (4.8,1.5) -- (2.9,1.5);
\node[teal] at (3.8,1.7){\small{$A^+$}};
\draw[red, thick] (4.8,1.4) -- (2.9,1.4);
\node[red] at (3.8,1.23){\small{$A^-$}};
\end{scope}

\draw[-,dotted,blue] (-0.6,1.4) to[bend left=30] (-0.6,0.6);
\draw[-,dotted,blue] (-0.6,0.6) to[bend right=30] (-0.6,-0.2);
\draw[-,dotted,teal] (-0.6,-0.3) to[bend left=30] (-0.6,-1.1);
\draw[-,dotted,teal] (-0.6,-1.1) to[bend right=30] (-0.6,-1.9);
\draw[-,dotted,red] (0.2,-2) to[bend left=90] (-0.6,-2);
\draw[-,dotted,red] (0.2,1.5) to[bend right=90] (-0.6,1.5);
\draw[-,dotted,red] (0.2,1.5) to[bend left=20] (0.2,-0.35);
\draw[-,dotted,red] (0.2,-0.35) to[bend right=20] (0.2,-2);

\end{tikzpicture}
\caption{Schematic representation of the replica trick for a single-interval entangling region $A$ with Rényi index $n=3$. }
\label{fig-replica}
\end{figure}
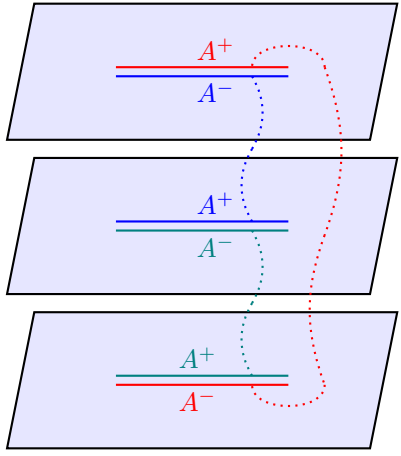

Before presenting the algebraic arguments, we first employ the replica method to argue that the vacuum $\Omega$ of QFT with local observables cannot have a flat entanglement spectrum. In this  setup, the moments of the density matrix  defining the R\'enyi entropies (\ref{renyi-entropies}) for a region $A$ can be computed as 
\be
\text{Tr}(\rho_{\Omega,A}^n) = \frac{Z[\mathcal{M}_{n, A}]}{(Z[\mathcal{M}])^n}\,,
\label{replica-trick}
\ee
where $Z[\mathcal{M}]$ is the partition function in the original manifold $\mathcal{M}$ and $Z[\mathcal{M}_{n,A}]$ is partition function evaluated on the $n$-sheeted replica manifold $\mathcal{M}_{n,A}$. This geometry is obtained by taking $n$ copies of the original space and sewing them together along an entangling region $A$, as depicted in Figure \ref{fig-replica}. For connected regions $A$, this process defines a manifold $\mathcal{M}_{n,A}$ with a codimension-two conical defect with opening angle $2\pi n$ (See \cite{Casini:2009sr,Rangamani:2016dms} for reviews on the topic). The flatness condition for the vacuum state would imply that the partition function of the replicated manifold has the following scaling 
\be 
Z[\mathcal{M}_{n,A}]= \frac{(Z[\mathcal{M}])^n}{\lambda_A^{n-1}} \label{scaling-replica}\,,
\ee
where $\lambda_A$ is a number defined through (\ref{renyi-entropies}) and should be understood as being finite only after introducing a regulator. However, the scaling (\ref{scaling-replica}) is not possible in a local QFT because the local degrees of freedom are sensitive to the presence of the conical singularity.  

Interestingly, the required scaling in (\ref{scaling-replica}) appears in topological field theories (TQFT) in the deep infrared at zero correlation length. In fact, in some examples, this can be recovered using \textit{surgery} methods \cite{Dong:2008ft} to obtain 
\be
Z[\mathcal{M}_{n,A}]=(Z[\mathcal{M}_1])^n \mathcal{D}^{n-1} \label{TFT-Z}\,,
\ee
where $\mathcal{D}$ is the total quantum dimension of the category defining the TQFT. Therefore, the vacuum of a TQFT can indeed be a stabilizer state, as already established in some specific discrete realizations~\cite{korbany2026long}, a paradigmatic one being the toric code~\cite{kitaev1997quantum}. In these realizations, the result generically includes an area term and the topological term $-\log\mathcal{D}$ \cite{Kitaev:2005dm}. In the presence of both terms, the entanglement spectrum can still be flat. In fact, this is the proper way to obtain a positive entropy, unlike the result of (\ref{TFT-Z}), which should only be understood as a scaling of the partition function.

To understand more precisely whether the flat-spectrum condition for stabilizer states can hold for the vacuum state $\Omega$ of a QFT in $D$ spacetime dimensions, we first note that the flatness condition holds for every spatial bipartition $A$ and $B$. We consider the entropies of the right wedge, namely the causal domain of the half-space $A$:
\be 
A=\big\{ x \in \mathbb{R}^{D-1,1} / x^0= 0, x^1\geq 0 \big\}\,. \label{right-wedge}
\ee
For such a region and the vacuum state, the Bisognano–Wichmann theorem \cite{BisognanoWichmann} fixes via Lorentz invariance the form of the density matrix to be
\be 
\rho_{\Omega,A}=\frac{e^{-2\pi K_{\Omega,A}}}{\Tr\big[e^{-2\pi K_{\Omega,A}}\big]}\,, 
\label{density-BW-matrix}
\ee
where $K_{\Omega,A}$ is known as the vacuum modular (or entanglement) Hamiltonian. On the $x^0=0$ Cauchy slice, it can be obtained via
\be
K_{\Omega,A}=\int_A d^{D-1}x \,x_1\,\mathcal{H}(x)
\label{modular-BW-hamiltonian}\,,
\ee
where $\mathcal{H}(x)$  is the energy density defined as the $T_{00}(x)$ component of the stress energy tensor of the QFT.  Equation (\ref{density-BW-matrix}) is not rigorous from an algebraic perspective as the reduced density matrix does not belong to the algebra of bounded operators in $A$. We return to this in section \ref{sec-2B}, where we also do not require the assumption of a local stress energy tensor. For now, we focus on the fact that, when discretized on a lattice with the proper cutoffs, the algebras of a QFT are of type I, and stabilizer states can be defined as in section \ref{sec-1A}. However, we will see that the entanglement spectrum recovered after regularizing the modular Hamiltonian, which represents a good approximation to leading order \cite{Giudici:2018izb,pourjafarabadi2021entanglement}, is not compatible with such a stabilizer structure. We begin by writing (\ref{modular-BW-hamiltonian}) on a lattice $\Lambda$ in the following way
\be 
K_{\Omega_\Lambda,A}=\sum_{X\subset A} x_{X,\Lambda}\mathcal{H}_{X,\Lambda}\,,
\label{modular-lattice-hamiltonian}
\ee
where $X$ denotes different sets of neighboring points contained in the lattice discretization of $A$, $x_{X,\Lambda}$ is the appropriate discretization of $x_1$ in the lattice $\Lambda$, and  $\mathcal{H}_{X,\Lambda}$ is the regularized local energy density that can be used to recover the lattice Hamiltonian $H_{\Lambda}$ via
\be 
H_{\Lambda}=\sum_{X\subset \Lambda}\mathcal{H}_{X,\Lambda}\,.
\label{hamiltonian-lattice}
\ee
The vacuum, even on the lattice, is a faithful state. In particular,  all eigenvalues of $\rho_{\Omega, A}$ in  (\ref{density-BW-matrix})  are strictly positive. If the vacuum were to have a flat entanglement spectrum, all eigenvalues would have to be non-trivial and equal. This condition extends to the modular Hamiltonian (\ref{modular-BW-hamiltonian}) as $K_{\Omega_\Lambda,A}=\lambda_{\Omega_\Lambda,A}\mathbb{I}_A$. Without loss of generality, we can decompose the energy density as $\mathcal{H}_X=E_{X,\Lambda} \mathbb{I}_A+\tilde{\mathcal{H}}_{X,\Lambda}$ where $\mathcal{H}_X$ should be understood as an operator in the full support of $\mathcal{H}_A$. Therefore, the requirement of flat spectrum implies
\bea 
&{}&\sum_{X\subset A}x_{X,\Lambda} E_{X,\Lambda}  = \lambda_{\Omega_\Lambda,A}\,,
\label{xH-lambda} \\
&{}&\sum_{X\subset A}x_{X,\Lambda} \tilde{\mathcal{H}}_{X,\Lambda}  = 0\,.
\label{xH-zero}
\eea
If we focus on the second equality (\ref{xH-zero}), constraining the non-diagonal contribution, we can expand $\tilde{\mathcal{H}}_{X,\Lambda}$ in some orthonormal basis for $\mathcal{B}(\mathcal{H}_\Lambda)$. For a qudit lattice, this could be the one of the generalized Pauli strings (\ref{pauli-string})
\be 
\tilde{\mathcal{H}}_{X,\Lambda}=\sum_{\textbf{a},\textbf{b},k} c^X_{\textbf{a},\textbf{b},k} \,\mathcal{P}_{\textbf{a},\textbf{b},k} \,.
\ee
where $c^X_{\textbf{a},\textbf{b},k}$ are, in principle, some arbitrary coefficients. However, the support of each Pauli strings is only non-trivial for some $X$ at fixed $x_{X,\Lambda}$. Together with linear independence, this leads to $ c^X_{\textbf{a},\textbf{b},k} =0$ and $\tilde{\mathcal{H}}_X=0$.  This highly constrains the Hamiltonian, as replacing it in (\ref{hamiltonian-lattice}) leads to
\be 
H_{\Lambda}=\Big(\sum_{X\subset A}E_{X,\Lambda}\Big)\mathbb{I}_A\oplus\Big(\sum_{X\subset B}E_{X,\Lambda}\Big)\mathbb{I}_B\,,
\ee
where we have applied the same arguments in the left wedge $B$ complementary to $A$. Considering reflection symmetry for the $x^1=0$ plane and translation invariance, we conclude that the Hamiltonian is simply proportional to the identity.

In conclusion, a Lorentz invariant QFT  governed by a non-trivial Hamiltonian in the regularized version can never yield a completely flat entanglement spectrum for the vacuum state.  This conclusion is borne out in known explicit models. See, for example,  \cite{Giudici:2018izb,pourjafarabadi2021entanglement} for numerical calculations in two-dimensional conformal field theories (CFT$_2$) with central charge $c=1/2,4/5,1$.

\subsection{III\texorpdfstring{$_1$}{1} factors as the root of magic in QFT  \label{sec-2B} }
In QFT, the algebras attached to regions are generically of type III$_1$ \cite{Buchholz:1986bg,Fredenhagen:1984dc} (See \cite{Yngvason:2004uh,WittenRev} for reviews). This implies that they are not isomorphic to the space of bounded operators $\mathcal{B}(\mathcal{H})$ in a given Hilbert space as in the type I case. Formally, this makes it difficult to define the corresponding reduced density matrices in the absence of a cutoff.

In type III factors, the reduced density matrix is replaced by the modular Tomita operator $\Delta_\psi$. But even the nature of this operator is different from the one we saw in type I algebras in (\ref{modular-operator}). In particular, the modular operator does not belong to $\alg$ but generates an outer automorphism acting as (\ref{modular-flow}). Nonetheless, it is uniquely defined in Tomita-Takesaki theory \cite{Takesaki:1970aki} via the spectral decomposition
\be 
\text{S}_\psi=\text{J}_\psi \Delta_\psi^{1/2}\,,
\ee
where S$_\psi$ is the Tomita operator defined by the action S$_\psi\,a\ket{\psi}=a^\dagger\ket{\psi}$ for all $a\in\alg$, and J$_\psi$ is an antiunitary operator mapping the algebra $\alg$ onto its commutant $\text{J}_\psi \alg \text{J}_\psi = \alg'$.
 
 To classify von Neumann factors from Tomita-Takesaki theory, Connes introduced the S$(\alg)$ invariant of the algebra $\alg$  \cite{Connes1973}. More precisely,  S$(\alg)$  is defined as the intersection of all spectra of the modular operator associated with faithful normal states allowed on $\alg$, or mathematically
 \be 
\text{S}(\alg) \;=\; \bigcap_{\substack{\psi \ \text{faithful} \\ \& \text{ normal}}} \mathrm{Spectrum}(\Delta_\psi)\,,
\label{connes} \ee
 where a state defined by a Hilbert space vector $\ket{\psi}$ on a von Neumann algebra $\alg$ is by definition normal, and it is faithful if
$\bra{\psi}a^\dagger a\ket{\psi}= 0$ implies $a=0$ for all $a \in \alg$.

The Connes invariant is continuous only for factors which are Type III$_1$. More precisely, 
 $S(\alg) = \{0\} \cup  \mathbb{R}^+$ for type III$_1$,  while, for example,  $S(\alg) = \{1\}$ for type I. This is the reason why stabilizer states with modular operators proportional to the identity, as the ones defined in section \ref{sec-1B}, are allowed in the lattice, but cannot be associated with faithful normal states in QFT. In other words, these faithful and normal states in QFT cannot exhibit the flat spectrum that characterizes stabilizer states. 

A natural question is how restrictive this faithful condition is. We begin by highlighting that every cyclic and separating vector defines, via its expectation value, a faithful normal state over the algebra. As a reminder, a state $\psi\in\mathcal{H}$ is said to be cyclic  with respect to an algebra  $\alg$ if $\{a \ket{\psi},\,\, a\in\alg\}$ is dense in $\mathcal{H}$. In addition, is separating if $a \ket{\psi} = 0$ implies $a= 0$ for $a\in \alg$. Physically, the key point is that, according to the Reeh-Schlieder theorem, the QFT vacuum $\Omega$ falls into this class. In other words, the vacuum is a cyclic and separating vector for the local algebra associated with any region. Furthermore, as the theorem holds for every region, every state that has sufficiently small distances approximates the vacuum should also be cyclic, separating, and therefore faithful and normal. This includes almost all physical states in a QFT, for example,  particle states, conformal primaries, and their descendants, etc.

From the point of view of modular theory, this is not surprising. In particular, for wedge regions, the action of the vacuum modular operator on the algebra, by Bisognano–Wichmann theorem \cite{BisognanoWichmann}, is described from  $\Delta_\Omega = e^{-2\pi K}$ with $K$ the Lorentz boost operator. $K$ has a continuous spectrum consisting of all real numbers, and therefore the spectrum of $\Delta_\Omega$ is also continuous and consists of the positive real numbers. For the remaining states, one can indeed check that their modular operator at sufficiently short distances can be approximated by the Bisognano–Wichmann one. This is done by using the scaling algebra formalism to study a given net of algebras at different energy scales \cite{Buchholz:1995gr,Buchholz:1997vu}. This fact is naturally useful to argue that for a cyclic and separating state in the continuum, the entanglement spectrum cannot become flat upon regularization following the proof in section \ref{sec-2A}.

Running the argument backwards, we conclude that every cyclic and separating state in a QFT with local observable operators must have a modular operator with a continuous spectrum and therefore cannot reproduce the characteristic flat entanglement spectrum with a trivial modular flow that characterizes stabilizer states.

Finally, to build intuition on why the separating condition is a source of magic, let us consider again a lattice regularization. Let us assume that exists a set of Pauli strings  $\mathbb{S}_{\psi_\Lambda}$ which stabilizes a physical state ${\psi_\Lambda}$ on the lattice $\Lambda$: $s_{\Lambda} \ket{\psi_\Lambda} = \ket{\psi_\Lambda}$ for all $s_{\Lambda}\in \mathbb{S}_{\psi_\Lambda}$. This implies that the regularized state is annihilated by the complementary operators 
\be 
(\mathbb{I}_\Lambda - s_{\Lambda}) \ket{\psi_\Lambda} = 0\,,\quad \forall s_{\Lambda}\in \mathbb{S}_{\psi_\Lambda}\,. \label{cyclic}
\ee
While taking the continuum limit with subsequent lattices, it may happen that all strings $s_{\Lambda}\in \mathcal{S}_{\psi_\Lambda}$ stop stabilizing the regularized vacuum at some given energy scale defining $\Lambda$, and  (\ref{cyclic}) is no longer valid. In this case, the state on the continuum will not originate from a stabilizer state. On the other hand, if (\ref{cyclic}) continues to hold in the continuum, it would imply that  $\mathbb{I}_\Lambda - s_{\Lambda}$ converges to an operator that annihilates the $\psi$. From the original definition (\ref{pauli-string}), a generic generalized Pauli operator will act non trivially in a bounded region. This region will have an associated type III$_1$ algebra $\alg$  for which $\psi$ is cyclic and separating.  Therefore, the only local operator capable of annihilating the $\psi$ is the trivial zero operator. This forces all  strings that stabilize the separating $\psi$ and survive in the continuum limit to converge to the identity.

\section{Examples and measures of Rényi flatness in QFT \label{sec-3}}
In this section, we provide examples of entanglement spectra for relevant states in QFT and show explicitly that these are not flat using different quantum information tools. We begin, in section \ref{sec-3A}, with standard calculations of a conformal field theory (CFT) vacuum Rényi entropies and mutual informations, including the $n\to0$ analytic continuation.  We also discuss excited particle states in the case of free boson in \ref{sec-3B}.

\subsection{Vacuum mutual information in CFT\label{sec-3A}}

We illustrate our result by showing that the vacuum of a CFT is always magical in any dimension. To begin, we consider the simplest example of CFT in $D=2$ space-time dimensions (CFT$_2$).  In this context,  non-stabilizerness was originally highlighted in \cite{Swingle:2020zoz} by numerical calculation in the critical Potts model and also observed in the Ising CFT$_2$ \cite{Hoshino:2025jko,Hoshino:2025ine} via computation of the Stabilizer Rényi entropies \cite{Leone:2021rzd} through the insertion of defects in the path integral. 

Our approach will be to simply put together well-known facts about the entanglement of the vacuum state. To begin, in a generic CFT$_2$, the Rényi entropies associated with one interval region $A_1=[u_1,v_1]$ of length $2r_1=v_1-u_1$ are famously \cite{Calabrese:2009qy} given by 
\be
 \Tr\big(\rho_{A_1}^n\big)
 \,=\, 
 C_n \,
 \Big|\,
 \frac{2r_1}{\epsilon} 
 \,\Big|^{-\tfrac{c}{6} (n -\frac{1}{n})}
 \,,
 \label{Fn1}
 \ee
 where $\epsilon$ is the UV cutoff and $C_n$ is a theory-dependent constant that comes from the normalization of twist operator correlation functions. This allows to compute the quantity
 \be 
 S_{n+1}(A_1)-S_n(A_1)=\frac{c\,\log({r_1}/{\epsilon})}{6\, n(n+1)} +\log \Big(\frac{C^{1/n}_{n+1}}{C^{1/(n-1)}_{n}}  \Big)\,.
\ee
This dependence on $n$ in the presence of a regulator can be considered an alternative argument that the vacuum of a CFT$_2$ is always magical, as (\ref{Fn1}) is independent of the operator content of the theory. A priori, the necessity of the dependence of the regulator seems annoying; however, a stabilizer state should show a flat spectrum for all values of $n$ and $r_1$ at a given energy scale, and this cannot be achieved by any proper choice of regulator $\epsilon$.

\begin{figure}[t]
\centering
\begin{tikzpicture}[
    scale=0.95,
    >=Latex,
    line cap=round,
    line join=round,
    every node/.style={font=\small}
]

\draw[very thick,->] (-4,0) -- (4,0) node[right] {$x$};
\draw[very thick,->] (-4,-2) -- (-4,2) node[above] {$t$};

\fill[blue!10]
    (-3,0) -- (-2,1) -- (-1,0) -- (-2,-1) -- cycle;
\draw[very thick,blue!70!black]
    (-3,0) -- (-2,1) -- (-1,0) -- (-2,-1) -- cycle;
\draw[ultra thick,blue!70!black] (-3,0) -- (-1,0);
\node[blue!70!black,below] at (-2,0) {$A_1$};
\fill[black] (-2.95,0) circle (2pt);
\node[black,above] at (-2.95,0.15) {$u_1$};
\fill[black] (-1.05,0) circle (2pt);
\node[black,above] at (-1.05,0.15) {$v_1$};

\fill[red!10]
    (0.8,0) -- (2.1,1.3) -- (3.4,0)-- (2.1,-1.3)-- cycle;
\draw[very thick,red!70!black]
    (0.8,0) -- (2.1,1.3) -- (3.4,0)-- (2.1,-1.3)-- cycle;
\coordinate (B1) at (0.8,0);
\coordinate (B2) at (3.4,0);
\draw[ultra thick,red!70!black] (0.8,0) -- (B2);
\node[red!70!black,below] at (2.1,0) {$A_2$};
\fill[black] (0.85,0) circle (2pt);
\node[black,above] at (0.85,0.15) {$u_2$};
\fill[black] (3.35,0) circle (2pt);
\node[black,above] at (3.35,0.15) {$v_2$};

\draw[thick,black] (-2,-1.5) -- (-1,-1.5);
\draw[thick,black] (-2,-1.45) -- (-2,-1.55);
\draw[dotted,black] (-2,-1) -- (-2,-1.45);
\draw[thick,black] (-1,-1.45) -- (-1,-1.55);
\draw[dotted,black] (-1,-0) -- (-1,-1.45);
\draw[thick,black] (0.8,-1.5) -- (2.1,-1.5);
\draw[thick,black] (0.8,-1.45) -- (0.8,-1.55);
\draw[dotted,black] (0.8,0) -- (0.8,-1.45);
\draw[thick,black] (2.1,-1.45) -- (2.1,-1.55);
\draw[dotted,black] (2.1,-1.3) -- (2.1,-1.45);
\draw[thick,black] (-2,1.5) -- (2.1,1.5);
\draw[thick,black] (-2,1.45) -- (-2,1.55);
\draw[dotted,black] (-2,1) -- (-2,1.45);
\draw[thick,black] (2.1,1.45) -- (2.1,1.55);
\draw[dotted,black] (2.1,1.3) -- (2.1,1.45);
\node[black] at (-1.5,-1.7) {$r_1$};
\node[black] at (1.4,-1.7) {$r_2$};
\node[black] at (-0.4,1.7) {$l$};

\end{tikzpicture}
\caption{Causal domain of two disjoint intervals $[u_1,v_1]$ and $[u_2,v_2]$ of length $2r_1=v_1-u_1$ and $2r_2=v_2-u_2$ respectively in $D=2$ space-time dimensions with separation $l$. }
\label{fig-intervals}
\end{figure}
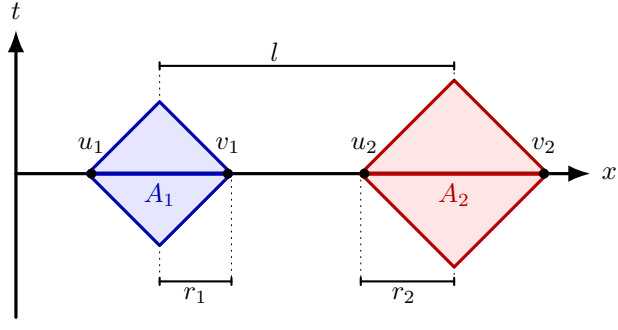

We can eliminate the regulator dependence altogether by considering the Rényi mutual information (\ref{mutual-information-definition}) between $A_1$ and another interval $A_2=[u_2,v_2]$ of length $2r_2=v_2-u_2$ at a distance $l$ from $A_1$. See figure \ref{fig-intervals} for a schematic depiction. As argued in section \ref{sec-1B}, this mutual information should also be independant on $n$ for a stabilizer state. In practice, the last term of (\ref{mutual-information-definition}) can be computed following \cite{Calabrese:2009ez,Calabrese:2010he} to obtain
\be
 \Tr\Big(\rho_{A_1 \cup A_2}^n\Big)
 =
 C_n^2 
 \Big|\frac{
 4r_1 r_2 (1-\x)}{\epsilon^2 }
 \Big|^{-\tfrac{c}{6} (n -\frac{1}{n})}
 F_n(\x)\,,
 \label{Fn}
 \ee
 with $\x$ the conformal cross ratio defined as
 \be
\x=\frac{(v_1-u_1)(v_2-u_2)}{(v_2-v_1)(u_2-u_1)}=\frac{4r_1 r_2}{l^2-(r_2-r_1)^2}\,,\label{crossratio}
 \ee
and $F_n(\x)$ a theory-dependent function that depends solely on the conformal cross ratio $\x$.
Using the replica method (\ref{replica-trick}), $F_n(\x)$ can be obtained from the genus $n-1$ partition functions on a Riemann surface with $\mathbb{Z}_n$ symmetry with period matrix
\be
{\tau}_{i,j}= \frac{2\textrm{i}}{n}\sum_{k=1}^{n-1}\sin\left(\frac{\pi k}{n}\right) \frac{ {}_2F_1(k/n,1-k/n; 1;1-\x )}{{}_2F_1(k/n,1-k/n; 1;\x) } \cos\left(\frac{2\pi k(i-j)}{n}\right)\,,
\label{period-matrix}
\ee
where ${}_2F_1$ represents the hypergeometric function. See \cite{GravaAbel,Headrick:2012fk,Coser:2013qda,Benedetti:2026nlt} for more extended discussions on the period matrix for different numbers of intervals and values of $n$. In this setup, if the vacuum has a flat spectrum as in (\ref{scaling-replica}), then it would imply that the genus $n-1$ partition functions can be obtained by elevating the partition function of the cylinder to the $n$th power. This is clearly not the case in a non-trivial CFT.

Combining (\ref{Fn1}) and (\ref{Fn}) in (\ref{mutual-information-definition}) yields a result independent of the cut-off, solely a function of the conformal cross-ratio $\x$:
\be
I_n(A_1,A_2)  = -\frac{c}{6} \left(1+ \frac{1}{n}\right)\log\left(1-\x\right) - \frac{\log F_n (\x)}{1-n}\,. 
\ee
This allows for the definition of regulator-independent measures of flatness. For example,  the difference of Rényi mutual informations 
\be
I_{n+1}(\x)-I_n(\x)= \frac{c \log(1-\x)}{6 n(n+1)} + \log\left(\frac{F_{n+1}(\x)^{1/n}}{F_n(\x)^{1/(n-1)}}\right) \label{mutual-information-difference}\,
\ee
is nonzero for generic $\x$, implying a nonflat spectrum.

We illustrate this result with a few examples. For $c=1$ theories we have \cite{Dijkgraaf:1987vp,Calabrese:2009ez} 
\be
F^{\text{c=1}}_{n}(\x) = \frac{1}{N(\x)^2}\left[\ThetaF{\pmb{0}}{\pmb{0}}(k\tau) \right]   \left[\ThetaF{\pmb{0}}{\pmb{0}}(\frac{\tau}{k}) \right]\,, \label{Fn-c=1}
\ee
where $k=1$ is the Wess-Zumino-Witten su$(2)_1$ theory, $k=2$ is the Dirac fermion and $k=4$ is the Kosterlitz-Thouless point. For the Ising model at $c=1/2$  we have \cite{Dijkgraaf:1987vp,Calabrese:2010he}
\be
F^{\text{Ising}}_{n}(\x)= \frac{1}{N(\x)}\sum_{\pmb{\alpha},\pmb{\beta}}\left|\ThetaF{\pmb{\alpha}}{\pmb{\beta}}(\tau) \right|\,, \label{Fn-Ising}
\ee
In both cases, the normalization is defined as in  \cite{Calabrese:2009ez,Calabrese:2010he} to ensure $F_{n}(0)=1$  so that the mutual information vanishes in the large separation limit, which yields
\be 
N(\x)={2^{n-1}}\left|\ThetaF{\pmb{0}}{\pmb{0}}(\tau) \right|\,.
\ee

\begin{figure}[t]
\centering
\includegraphics[width=1\textwidth]{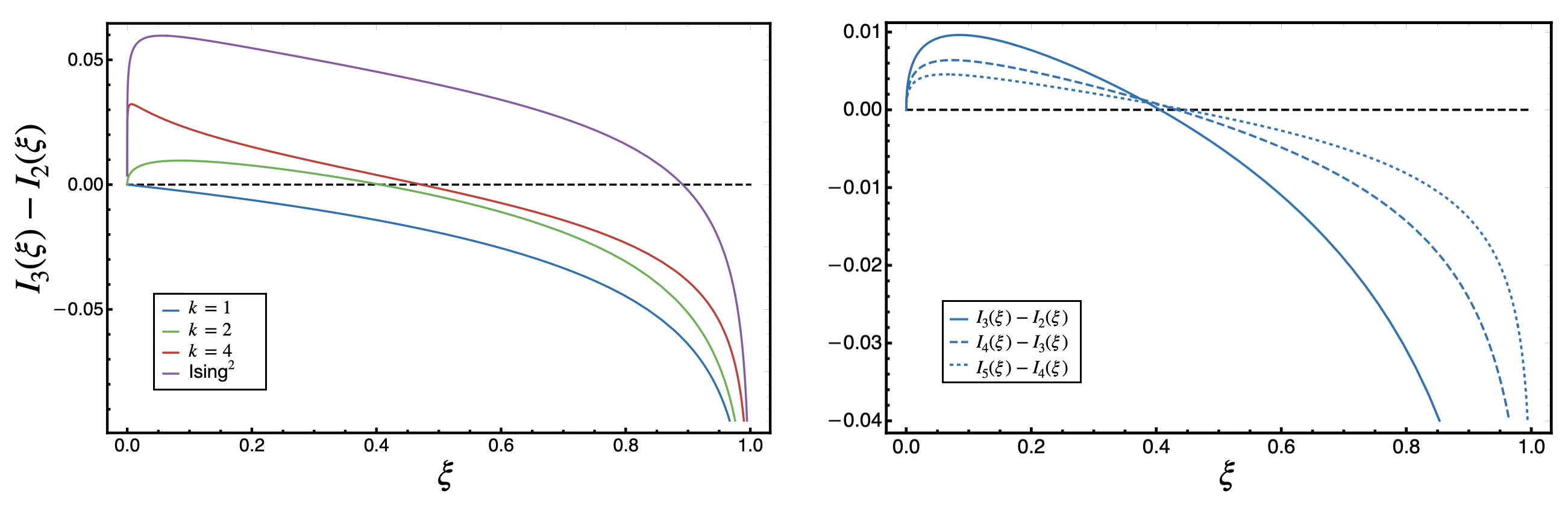}
\caption{Difference between mutual informations $I_{n-1}-I_n$ of two disjoint intervals as a function of the cross-ratio $\x$ for $n=2$ and different $c=1$ CFT$_2$ given by (\ref{Fn-c=1}) at $k=1,2,4$ and (\ref{Fn-Ising}) (left) and for $n=2,3,4$ in the $k=2$ case (right).} 
\label{fig-mutual-information}
\end{figure}

In figure \ref{fig-mutual-information}, we plot the difference in Rényi mutual information $I_3(\x)-I_2(\x)$, as a function of the cross ratio $\x$ defined in (\ref{crossratio}) for $c=1$ theories. The curves are obtained from (\ref{mutual-information-difference}) for $n=2$ using (\ref{Fn-c=1}) for $k=1,2,4$, together with the case of the orbifold theory corresponding to two copies of the Ising CFT$_2$ given by (\ref{Fn-Ising}). In figure \ref{fig-mutual-information}, we also show for $k=2$ the mutual information difference for $n=2,3,4$. We highlight that the flattening of the curve with increasing $n$ is expected, as for bigger values of $n$, effectively fewer eigenvalues of the density matrix contribute. We note that, although values of $\x$ exist for which $I_{n+1}(\x)=I_n(\x)$, this has no implications for stabilizerness, as the flatness condition should hold for all regions; therefore, independently of $\x$.

The mutual information of disjoint regions is also finite in general $D$ space-time dimensions. For convenience, we consider the case for two disjoint spheres $A_1$ and $A_2$ of radius $r_1$ and $r_2$ respectively separated by a distance $l$. These obey the expansion \cite{Cardy:2013nua} 
\be 
I_n(A_1, A_2) = \sum_{\phi_j}C_{A_1}^{(n)}({\phi_j})C_{A_2}^{(n)}({\phi_j})\left( \frac{r_1 r_2}{4l^2} \right)^{\sum_{j=1}^{n} \Delta_{\phi_j}}\,. \label{conformal-expansion}
\ee
where the sum should be understood over the primaries ${\phi_j}$  of scaling dimensions $\Delta_{\phi_j}$. In particular, at very long separation $r_1 r_2\ll l^2$, the mutual decay is given by the scaling dimension of the lowest scaling primary field.  For example, in the free scalar CFT in $D=4$ space-time dimensions, the long-distance leading contribution takes the form
\be
    I_n(A_1, A_2) \sim \frac{n^4-1}{15n^3(n-1)} \left( \frac{r_1 r_2}{l^2} \right)^{2}\,,
\ee
and therefore we have that
\be
    I_{n+1}(A_1, A_2)- I_n(A_1, A_2) \sim-\Big(\frac{n^4+4n^3+7n^2+4n+1}{15n^3(n+1)^3} \Big)\left( \frac{r_1 r_2}{l^2} \right)^{2}\,.
\ee
This does not vanish for any value of $n$. Therefore, it implies that the vacuum of a free scalar CFT in $D=4$ dimensions cannot be flat and is indeed magical. A similar expansion can be found for other more general CFTs \cite{Casini:2021raa}\footnote{We add that the expansion (\ref{conformal-expansion}) of the mutual formation can be computed exactly for a general CFT and for arbitrary $l$ using the structure of the modular flow \cite{Agon:2024xvs, Agon:2026jjh} and the conformal block expansion \cite{Chen:2017hbk}.}.

Another convenient probe to show that the vacuum of a CFT is generically non-flat is the analytic continuation of Rényi entropies in the $n\to 0$ limit. Even though this quantity is regulator dependent as in (\ref{Fn1}), it contains universal information. Moreover, it has the advantage that, for a spherical region $A$ of radius $R$, the limiting behavior as $n\to 0$ can be easily computed for a generic CFT. 

For the calculation, we highlight that the modular Hamiltonian of the sphere can be obtained by conformally mapping the Bisognano-Wichmann modular Hamiltonian (\ref{modular-BW-hamiltonian}) using the conformal transformation \cite{Casini:2011kv}
\be
x^\mu\to \frac{x^\mu-x^2 c^\mu}{1-2 (x\cdot c)+ (x\cdot x)(c\cdot c)}+2 Rc^\mu \,, \quad c^\mu=(0,1/2R,0,\dots,0)\,.
\ee
As a result, the $n$th power of the density matrix can be obtained as
\be 
\rho_A^n=\frac{e^{-2\pi n K_{\Omega,A}}}{\Tr[e^{-2\pi n K_{\Omega,A}}]},\quad K_{\Omega,A}(r)=\int_{A} d^{D-1}x\left( \frac{R^2-r^2(x)}{2R}\right)\mathcal{H}(x)\,. 
\ee
In the $n\to0$ limit of the vacuum Rényi entropies can then be computed in analogy to a free energy of a thermal bath as from (\ref{replica-trick}) we get $S_n (A) \sim \log\big( Z[\mathcal{M}_n]\big)$. In particular, it corresponds to the free energy of a highly excited state with vanishing inverse temperature depending on $x$ as  
\be 
\beta_{eff}(x)= 2\pi n \left( \frac{R^2-r^2(x)}{2R}\right)\,.
\ee
In fact, it was proven in \cite{Agon:2023tdi} that it can be computed, introducing a cutoff $\epsilon$, via
\be
S_n(A)\sim\sigma\int_{A}\frac{d^{D-1}x}{\beta_{eff}(x)^{D-1}}\,= \frac{\sigma R^{D-1} S_{D-2}}{\pi^{D-1} n^{D-1} }\int_0^{R-\epsilon}\frac{r^{D-2} dr}{(R^2-r^2(x))^{D-1}}\,, \label{renyi0}
\ee
where $S_{D-2}={2 \pi^{(D-1)/2}}/{\Gamma[(D-1)/2]}$ is the surface of the $D-2$ sphere in terms of the gamma function $\Gamma$, and  $\sigma$  is known as the \textit{Boltzmann} constant. For a given CFT, $\sigma$ is defined  from the growth of the free energy $F$ with volume $V$ and temperature $1/\beta$ as $F=\sigma {V}{\beta^{-D}}$. In particular, for the free boson it is given by $ {\zeta[D]\Gamma[D/2]}{\pi^{-D/2}}$ where $\zeta$ is the Euler zeta function.

To proceed, it is convenient to express (\ref{renyi0}) using the volume of a ball of radius $\nu_{\text{max}}$ inside the hyperbolic space $\mathbb{H}^{D-1}$ as
\be
S_n(A)\sim\frac{\sigma  S_{D-2}}{(2\pi n)^{D-1} V_{D-2} }V[\mathbb{H}^{D-1},\nu_{\text{max}}]\,, \quad \nu_{\text{max}}=\tanh^{-1}\left(\frac{2(1-\epsilon/R)}{1+(1-\epsilon/R)^2}\right)\,,
\ee
where $V_{D-2}={\pi^{(D-2)/2}}/{\Gamma[D/2]}$ is the volume of the $D-2$ sphere. Also, we note that the dependence on the Rényi index $n$ is non-trivial and fixed by the coefficient. Furthermore, in general even dimensions, there is a universal logarithmic term which coefficient does not depend on the cutoff. It can be obtained, considering the logarithmic term in the expansion
\bea
V[\mathbb{H}^{D-1},\nu_{\text{max}}] &=& V_{D-2} \int_0^{\nu_{\text{max}}} d\nu \,\sinh^{D-2}(\nu)=\\ &=&\dots+ (-1)^{\frac{D-2}{2} }\frac{V_{D-2}\pi^{\frac{D-2}{2}}}{S_{D-2 }\Big(\frac{D-2}{2}\Big)!} \log\left(\frac{R}{\epsilon}\right)+\dots\,, \label{expand-volume}
\eea
which produces the universal term in the entropy
\be
S_n(A)\sim \dots+ \frac{\sigma}{n^{D-1}}\left(\frac{(-1)^{\frac{D-2}{2} }}{2^{d-1}\pi^{\frac{D}{2}}\Big(\frac{D-2}{2}\Big)!}\right)\log\left(\frac{R}{\epsilon}\right)+\dots\,.
\ee
We highlight that the fact that close to the $n\to0$ limit, there is a non trivial dependence of $n$ in the universal coefficients of the entropy for every CFT signals the fact that the vacuum of the CFT cannot correspond to a flat spectrum. These methods can be extended to odd $D$ by computing the universal constant term in the expansion of (\ref{expand-volume}) and to holographic CFT or more general QFT following \cite{Agon:2023tdi,Agon:2025wtq}. 

\subsection{Particle states in the free boson QFT \label{sec-3B} }

As the next example, we consider the entanglement spectrum associated with the density matrix in the right Rindler wedge (\ref{right-wedge}) for particle states of a free boson in Rindler spacetime.
For the Minkowski vacuum this provides a free-field realization of the Bisognano-Wichmann theorem,  but for a free field one can consider also more general Fock states and construct the reduced density matrix explicitly using the Bogoliubov coefficients. 

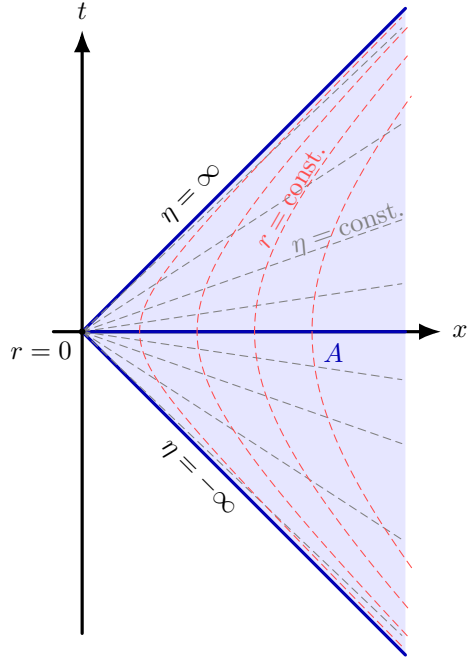
\begin{figure}[t]
\centering
\begin{tikzpicture}[
    x=1cm,y=1cm,
    line cap=round,
    line join=round,
    >=Latex,
    scale=0.95,
    every node/.style={font=\small}
]

\fill[blue!10] (0,0) -- (4.5,4.5) -- (4.5,-4.5) -- cycle;
\draw[very thick,->] (-0.4,0) -- (5,0) node[right] {$x$};
\draw[very thick,->] (0,-4.2) -- (0,4.2) node[above] {$t$};
\draw[very thick,blue!70!black] (0,0) -- (4.5,4.5);
\draw[very thick,blue!70!black] (0,0) -- (4.5,-4.5);
\draw[very thick,blue!70!black] (0,0) -- (4.5,0);
\foreach \m in {-0.95,-0.65,-0.35,-0.15,0.15,0.35,0.65,0.95} {
  \draw[densely dashed,gray!99] (0,0) -- (4.45,{4.45*\m});
}

  \draw[densely dashed,red!70]
    plot[domain=-2.4:2.4,samples=140,smooth]
    ({0.8*cosh(\x)},{0.8*sinh(\x)});
  \draw[densely dashed,red!70]
    plot[domain=-1.7:1.7,samples=140,smooth]
    ({1.6*cosh(\x)},{1.6*sinh(\x)});
\draw[densely dashed,red!70]
    plot[domain=-1.25:1.25,samples=140,smooth]
    ({2.4*cosh(\x)},{2.4*sinh(\x)});
 \draw[densely dashed,red!70]
    plot[domain=-0.9:0.9,samples=140,smooth]
    ({3.2*cosh(\x)},{3.2*sinh(\x)});

\fill (0,0) circle (1.3pt);
\node[below left] at (0,0) {$r=0$};

\node[rotate=45,anchor=south west] at (1.25,1.25) {$\eta=\infty$};
\node[rotate=-45,anchor=north west] at (1.25,-1.25) {$\eta=-\infty$};
\node[rotate=19,anchor=north west,gray!99] at (2.7,1.4) {$\eta=\text{const.}$};
\node[rotate=60,anchor=north west,red!70] at (2.2,1.2) {$r=\text{const.}$};

\node[ultra thick,blue!70!black]  at (3.5,-0.3) {$A$};

\end{tikzpicture}
\caption{Right Rindler wedge  $A$ in $D=2$ Minkowski space-time described by the Rindler coordinates $\eta\in(-\infty,\infty)$ and $r=[0,\infty)$. }
\label{fig-rindler}
\end{figure}

To ignore the transverse directions, we consider $D=2$ Minkowski spacetime with metric $ds^2= -dt^2+dx^2$ and introduce Rindler coordinates $r$ and $\eta$ by
\be 
t=r\sinh(a\eta),\quad x=r\cosh(a\eta) \,,
\label{coordinates-rindler}
\ee
where $a>0$ is a the acceleration parameter. For $\eta\in(-\infty,\infty)$ and $r=[0,\infty)$. These are the natural coordinates  for a uniformly accelerated observer for describing the right Rindler wedge (\ref{right-wedge}) with the metric
\be 
ds^2= -r^2 a^2 d\eta^2+dr^2\,.
\label{metric-rindler}
\ee
The lines of constant $r$ are the world lines of uniformly accelerated observers with proper acceleration $r$ and the constant $\eta$ straight lines are the Rindler time slices. The Rindler time translations correspond to Lorentz boosts.   See figure \ref{fig-rindler} and  also \cite{Birrell:1982ix} for a review.

Since the constant $r$ lines cannot cross the null rays at $\eta\to\pm\infty$, the null rays are the event horizon for the uniformly accelerated observer. From the perspective of observers in the right or left wedges, the natural vacua are $\ket{0}_R$ or  $\ket{0}_L$ which do not coincide with the Minkowski vacuum $\ket{0}_M$. The action of creation and annihilation operators in the left and right wedges given by  $b^\dagger_{L,\omega},b_{L,\omega}$ and $b^\dagger_{R,\omega}. b_{R,\omega}$ is totally decoupled. Such operators are defined as usual by their commutation relations or equivalently by their action on left $\ket{n}_L$ and right $\ket{n}_R$ particle states:
\bea 
b_{L,\omega} \ket{n}_L \ket{n}_L &=& \sqrt{n} \ket{n-1}_L \ket{n}_R \,,\label{b1}\\
b_{R,\omega} \ket{n}_L \ket{n}_R &=& \sqrt{n} \ket{n}_L \ket{n-1}_R\,,\label{b2} \\
b_{L,\omega}^\dagger \ket{n}_L \ket{n}_R &=& \sqrt{n+1} \ket{n+1}_L \ket{n}_R\,,\label{b3} \\
b_{R,\omega}^\dagger \ket{n}_L \ket{n}_R &=& \sqrt{n+1} \ket{n}_L \ket{n+1}_R\,.\label{b4} 
\eea

To obtain the reduced density matrix of  excited particle states, it is convenient to consider  the Unruh creation operators, which can be obtained from a Bogoliubov transformation \cite{Unruh:1976db,Marolf:2003sq,Marolf:2004et}  of the Rindler  creation and annihilation operators (\ref{b1}-\ref{b4}): 
\bea 
a_{(1),\omega}^\dagger&=& \frac{1}{\sqrt{1-q_\omega}}\Big(b_{R,\omega}^\dagger -q_\omega^{1/2} b_{L,\omega}\Big)\label{creation-1-rindler}\,, \\
a_{(2),\omega}^\dagger&=&\frac{1}{\sqrt{1-q_\omega}}\Big(b_{L,\omega}^\dagger -q_\omega^{1/2} b_{R,\omega}\Big)\label{creation-2-rindler} \,.
\eea
with $q_\omega$ depending on the frequency $\omega$ as
\be 
q_\omega=e^{-2\pi \omega/a}=e^{-\omega/T_{\text{U}}}\,.
\ee
where $T_{\text{U}}=a/2\pi$ is the Unruh temperature \cite{Unruh:1976db}. As is well known, the Minkowski vacuum is annihilated by the Unruh annihilation operators \cite{Unruh:1976db}. Therefore, 
\be 
a_{(1),\omega}\ket{0}_M=a_{(2),\omega}\ket{0}_M=0 \label{mink-vac}\,.
\ee
Combining (\ref{creation-1-rindler}), (\ref{creation-2-rindler}), and (\ref{mink-vac}), it is clear that the Minkowski vacuum can be viewed as a two-mode squeezed state of the left and right Rindler modes: 
\bea 
\ket{0}_M &=&\prod_\omega \sqrt{1-q_\omega} \exp \big[\sqrt{q_\omega} b_{L,\omega}^\dagger b_{R,\omega}^\dagger \big] \ket{0}_L \ket{0}_R  \nonumber \\ 
&=&\prod_\omega \sqrt{1-q_\omega} \sum_{n=0}^\infty q_\omega^{n/2} \ket{n}_L \ket{n}_R\,,
\label{vacuum-rindler}
\eea
The physical implication is that $\ket{0}_M$ can be expressed as an entangled state between the left and right wedges \cite{Unruh:1976db}.

Our aim in this section is to study particle states. We begin noticing that any state created by (\ref{creation-1-rindler}) and (\ref{creation-2-rindler}) acting on (\ref{vacuum-rindler}) can be associated with a density matrix of the form $\rho=\otimes_\omega \rho_\omega$. This implies that the entropy decomposes as 
\be 
S_n (\rho)=\sum_\omega S_n (\rho_\omega)\,.
\ee
Since $S_n (\rho_\omega)\leq S_{n'} (\rho_\omega)$ for all $n'<n$,  a flat spectrum requires $S_n (\rho_\omega)= S_{n'} (\rho_\omega)$ for all fixed values of $\omega$. To try to saturate this inequality, we consider states of the form 
\be
\ket{N}_{M,\omega}= \frac{(a_{(1),\omega}^\dagger)^{N}}{\sqrt{N!}}\ket{0}_{M,\omega} \label{unruh-states}\,,
\ee
which correspond to Minkowski particle states mainly localized in the right wedge. However, for $\omega\gg 1$ we have that $a^\dagger_{(1),\omega}$ is good approximation for the usual Minkowski creation operator \cite{Casini:2008cr}. The corresponding reduced density matrix for (\ref{unruh-states}) in the right wedge (\ref{right-wedge}) can be computed from  (\ref{vacuum-rindler}) and  (\ref{creation-1-rindler}) to be 
\bea 
 \rho_{N,\omega,R} &=& \Tr_L\Big(\ket{N}_{M,\omega}\bra{N}_{M,\omega}\Big) \\ &=& (1-q_\omega)^{N+1} \sum_{m=0}^\infty  q_\omega^{m}\frac{(m+N)!}{N! m!} \ket{m+N}_R \bra{m+N}_R \,. \nonumber
\eea
Furthermore, the trace  of $\rho_{N,\omega,R}^n$ can be written as a sum of powers of a Pascal distribution
\be 
P(f,s;p)=\frac{(s+f-1)!}{f-1!s!}(1-p)^f p^s 
\label{pascal-distribution}\,,
\ee
which is the probability density function for $f$ failures and $s$ successes with $p$ being the probability of success on each trial. In particular, we obtain
\bea 
 \Tr_R\Big(\rho_{N,\omega,R}^n\Big)  &=&  \sum_{m=0}^\infty\Big[  P(N+1,m,q_\omega) \Big]^n  \label{trace-computed}\\
 &=& (1-q_\omega)^{(N+1)n} {}_nF_{n-1} \big( \underbrace{N+1, \dots, N+1}_{n \text{ times}}; \underbrace{1, \dots, 1}_{n-1 \text{ times}}; q_\omega^n \big)\nonumber\,,
\eea
where we have also expressed the result in terms of the generalized hypergeometric function
\be 
{}_nF_m(\overline{a};\overline{b}; z) = \sum_{k=0}^{\infty} \frac{(a_1)_k (a_2)_k \dots (a_n)_k}{(b_1)_k (b_2)_k \dots (b_m)_k} \frac{z^k}{k!}\,,
\ee
for $\overline{a}=(a_1,\dots, a_n)$, $\overline{b}=(b_1,\dots, b_m)$ and 
\be
(a)_n =\left\{ \begin{array}{cc}
   0  & \text{if }n=0 \\
   (a +n-1)!/(a-1)! & \text{if }n>0
\end{array}\right.
\ee
This is more convenient for numerical manipulations. 

\begin{figure}[t]
\centering
\includegraphics[width=1.0\textwidth]{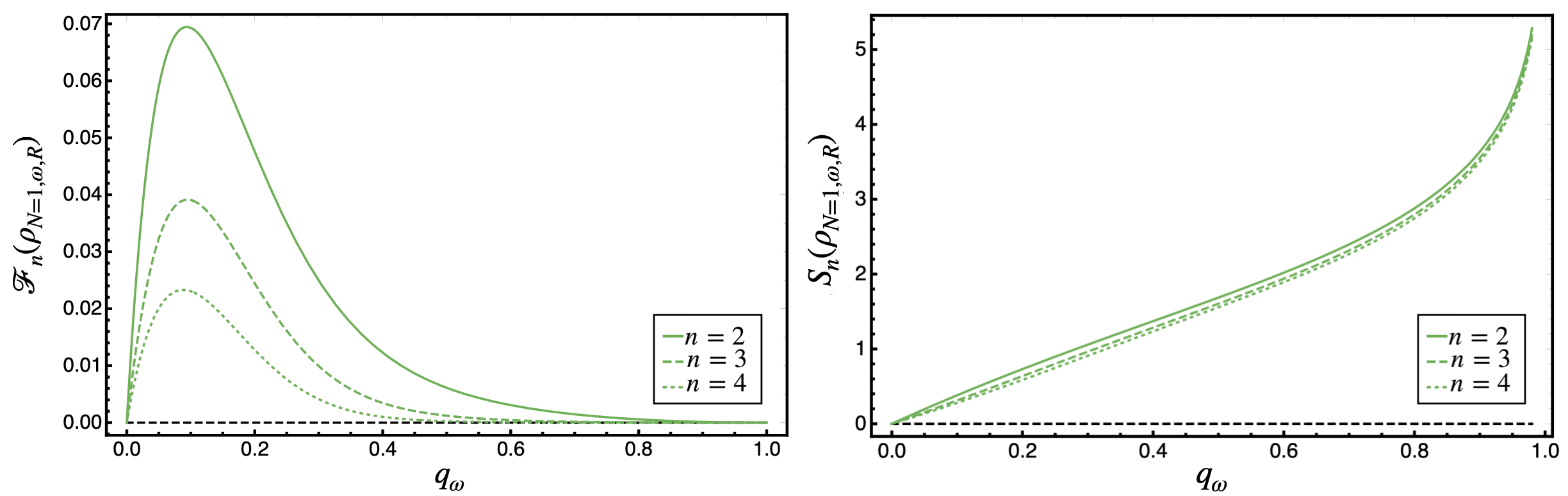}
\caption{Antiflatness (left) and Rényi entropies (right) as a function of $q_\omega$ for a one Unruh particle state $N=1$ and Rényi index $n=2,3,4$.}
  \label{figure-vary-renyi}
\end{figure}

A useful quantity to analyze the entanglement spectrum at fixed $\omega$ is the anti-flatness introduced in \cite{Tirrito:2023fnw} defined for arbitrary values of the Rényi index
\be 
\mathcal{F}_n(\rho_A)= \Tr_A(\rho_A^{n+1})- \Tr_A^{\frac{n}{n-1}}(\rho_A^n)\,.
\label{anti-flatness}
\ee
In figure \ref{figure-vary-renyi} we show for a one Unruh particle state, given by (\ref{unruh-states}) with $N=1$, the behavior of the anti-flatness $\mathcal{F}_n(\rho^R_{N=1,\omega})$ and Rényi entropies $\mathcal{S}_n(\rho^R_{N=1,\omega})$ for $n=2,3,4$ obtained from (\ref{trace-computed}). We can see a peak in the anti-flatness that vanishes for very high or low frequencies. However, not surprisingly, the peak decreases with $n$ due to the fact that at a larger Rényi index, fewer eigenvalues effectively contribute to the entropy. 

\begin{figure}[t]
\centering
  \includegraphics[width=1\textwidth]{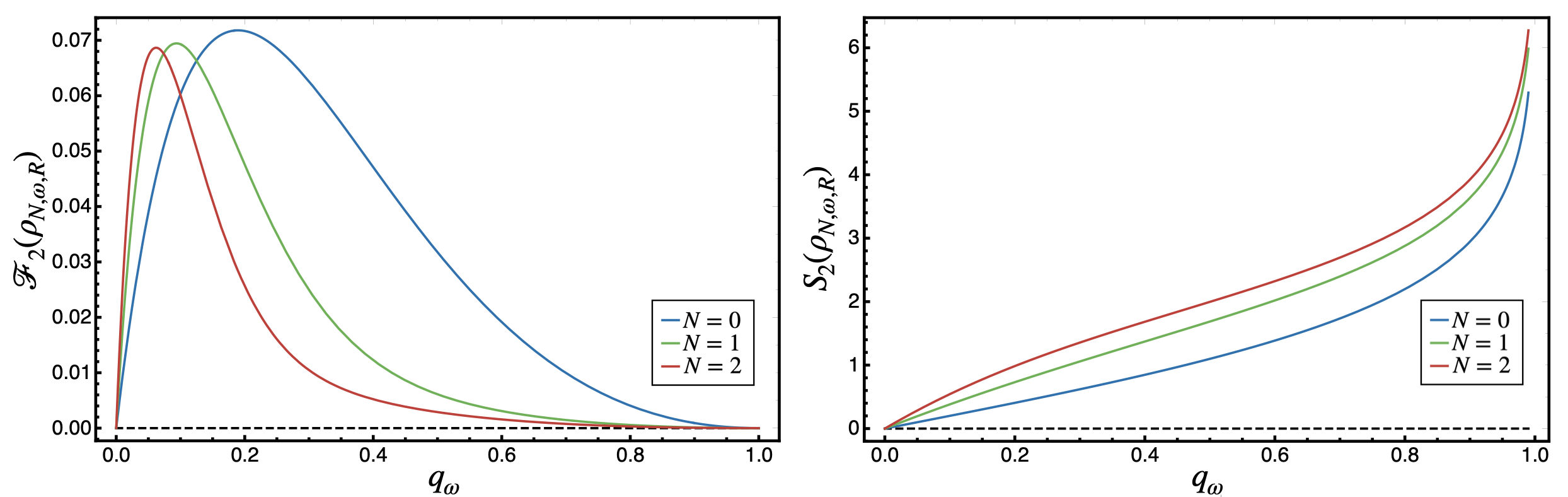}
   \caption{ Antiflatness (left) and Rényi entropies (right)  for fixed Rényi index $n=2$ as a function of $q_\omega$ for the vacuum and one and two Unruh particle states $N=0,1,2$. }
   \label{figure-vary-particle}
\end{figure}

In this regard, the quantity of main interest is $\mathcal{F}_2(\rho_A)$, which encodes more information about the entanglement spectrum than the counterparts for $n>2$. It was shown \cite{Tirrito:2023fnw} that after averaging over a Clifford orbit in a quantum system, $\mathcal{F}_2(\rho_A)$ is equivalent to the linearized stabilizer entropy. Such averaging is complicated to understand in QFT. Nonetheless, in \cite{Cao:2024nrx} it was shown that $\mathcal{F}_2(\rho)$ represents a lower bound on the magic of a finite quantum system of the form
\be 
\mathcal{F}_2(\rho_A)\leq 8\,\min_{\sigma \in \mathrm{STAB}} || \psi-\sigma||
\label{anti-flatness-bound}\,,
\ee
where STAB denotes the space of all stabilizer states. In figure \ref{figure-vary-particle}, we show the behavior of the anti-flatness $\mathcal{F}_2(\rho^R_{N,\omega})$ and Rényi entropies $\mathcal{S}_2(\rho^R_{N,\omega})$ for states defined by $N=0,1,2$ in (\ref{unruh-states}). Interestingly, while the entropy increases with particle number $N$, the spectrum becomes flatter. 

In the large particle limit,  the Pascal distribution (\ref{pascal-distribution},\ref{trace-computed}) can be approximated by a Gaussian:
\be
P(N+1, m, q_\omega) \approx \frac{1}{\sqrt{2\pi \sigma^2}} \exp\left( - \frac{(m - \mu)^2}{2\sigma^2} \right)\,,
\ee
with mean $\mu$ and variance $\sigma$ given by  
\be 
\mu = (N+1)\frac{q_\omega}{(1-q_\omega)}\,,\quad  \sigma^2 =(N+1)\frac{q_\omega}{(1-q_\omega)^2}\,.
\ee
This generates a non-flat spectrum by (\ref{trace-computed}) for all finite values of $N$. However, such spectra become flatter as $N\to \infty$. In the limit $N\to \infty$, we have that $P(N+1, m, q_\omega)\to0$ for finite values of $\omega$.  Therefore, in the $N\to \infty$ limit, the state (\ref{unruh-states}) seems to become flat. This state is, however, not formally included in the original Fock space and is a singular state from the algebraic point of view. 

\section{Conclusions}\label{conclusions}

We have shown that in quantum field theory, the entanglement spectrum of any physical state  - a state that resembles the vacuum at sufficiently short distances - cannot be flat. This is a simple consequence of the Bisognano-Wichmann theorem.  Another reasonable requirement for a physical state is that the expectation value of the renormalized energy-momentum tensor in this state should be finite, which would exclude artificial states like the Rindler or the Boulware vacuum from these considerations. Since entanglement flatness is a necessary condition for stabilizer-ness, our result shows that any physical state in a QFT necessarily has non-zero magic.

Our results also indicate that, within the framework of quantum error correction, the relation between computational complexity and field theory is quite distinct from that of other common quantum resources. In particular, the role of magic is fundamentally distinct from that of non-Gaussianity.  As a witness, magic delineates a clear boundary between TQFT and local QFT, rooted in their algebraic structure. On the other hand, non-Gaussianity may not be sensitive to this distinction but is capable of distinguishing free theories from interacting ones.  An interesting direction to pursue would be to understand how magic resources relate to sampling the path integral of a QFT on a lattice (for example,  in lattice QCD), as the latter is not, {\it a priori}, necessarily bound by neither magic nor entanglement. 

\newpage
\vspace{0.5cm}
Furthermore, our analysis has implications for gravitational theories. One expects that even in weakly coupled gravity, the vacuum will continue to be cyclic and separating in perturbative effective quantum field theory. For example,  the quantum state of fields near the horizon of an eternal black hole is expected to be the Hartle-Hawking vacuum and not the Boulware vacuum.  As a result, one expects a non-flat spectrum and therefore non-zero magic for such physical states in gravity. In a complete and finite theory of gravity like string theory, one can regard low energy gravity as an effective field theory with the string length effectively acting as a lattice cutoff. In this setup, our conclusions are still valid.

Finally, we comment briefly on the implications of our results for AdS/CFT holography. Using the Ryu-Takayanagi correspondence  \cite{Ryu:2006bv,Lewkowycz:2013nqa,Dong:2016fnf},   it is easy to construct states in the boundary with a flat spectrum using the gravitational path integral \cite{Dong:2018seb} in the bulk. These are known as fixed-area states, and their leading bipartite entanglement spectrum coincides with the spectrum of stabilizer states \cite{Chemissany:2025vye, Nezami:2016zni, Akers:2019gcv, Mori:2024gwe, Li:2025nxv} from the perspective of tensor networks \cite{QEC1, QEC2,QEC3}.  However, from a boundary perspective, the CFT should be described as a net of type III$_1$ algebras associated to sub-regions\footnote{This is clear in the $N\to\infty$ limit, when the theory is described by a net of Generalized Free Fields, as they satisfy the local QFT axioms \cite{Greenberg:1961mr}. See also \cite{Duetsch:2002hc,Benedetti:2022aiw,Leutheusser:2022bgi} for more recent holographic discussions in the context of subalgebra-subregion duality.}. This fact implies that there cannot be faithful states with a flat spectrum and, therefore, faithful stabilizer states. Also, in the bulk we expect physical states even in the presence of weak gravity to be cyclic and separating, as argued in the previous paragraph. In principle, these states can be written in the basis of fixed area states. In this case, one obtains a non-flat sub-leading contribution to the Rényi entropies using the modified cosmic-brane proposal \cite{Dong:2023bfy,Penington:2024jmt}. This is consistent with recent results \cite{Cao:2024nrx,Cao:2026uoq}. 

\vspace{0.8cm}
\acknowledgments
\vspace{-0.1cm}
We are grateful to Hernan Bueno Xavier, Horacio Casini, Wissam Chemissany, Marina Huerta, and Emanuele Tirrito for insightful discussions. We also thank  Horacio Casini and Wissam Chemissany for valuable comments on previous versions of this manuscript.  V.B. acknowledges the support of a Research Fellowship from the Abdus Salam International Centre for Theoretical Physics (ICTP), Trieste, Italy, and by INFN Iniziativa Specifica ST$\&$FI. M.~D. was partly supported by the EU-Flagship programme Pasquans2, by the PNRR MUR project PE0000023-NQSTI, by INFN Iniziativa Specifica Quantum, and by the ERC Consolidator grant WaveNets (Grant agreement ID: 101087692). 

\newpage
\bibliographystyle{JHEP}
\bibliography{biblio}

\end{document}